\documentclass[a4paper,12pt]{article}
\begin{document}
\title{Update on multiparticle effects in Bose-Einstein correlations}
\author{Kacper Zalewski\thanks{Supported in part by the KBN grant
2P03B 093 22}\\
M. Smoluchowski Institute of Physics, Jagellonian University\\ and\\ Institute
of Nuclear Physics, Krak\'ow, Poland} \maketitle
\begin{abstract}
Multiparticle effects in Bose-Einstein correlations are reviewed.
It is shown that for a broad class of models they can be ignored
in the low density limit, but often are significant (typically at
a $10$\% level) for realistic final states.
\end{abstract}
Bose-Einstein correlations.\vspace{1cm}

\section{Statement of the problem}

Most models of the Bose-Einstein correlations start with an input
single particle density matrix $\rho_I(\textbf{p};\textbf{p}')$ or
some equivalent information. The subscript $I$ reminds that this
is an input and not necessarily the actual single particle density
matrix of the system being considered. This input depends on the
quantities of interest e.g. on the radii of the interaction
region. Let us concentrate on a system of $n$ identical bosons. If
these particles were distinguishable, one could build a reasonable
$n$-particle density matrix as a product of the single particle
density matrices $\rho_I$:

\begin{equation}\label{}
  \rho^U_n(\textbf{p}_1,\ldots,\textbf{p}'_n) = \prod_{j=1}^n \rho_I(\textbf{p}_j;\textbf{p}'_j).
\end{equation}
For identical bosons, however, this product must at least be
symmetrized \cite{KAR}, \cite{BIK} and one finds

\begin{equation}\label{}
  \tilde{\rho}_n(\textbf{p}_1,\ldots,\textbf{p}'_n) = \sum_P \prod_{j=1}^n
  \rho_I(\textbf{p}_i;\textbf{p}'_{Pi}),
\end{equation}
where the summation is over all the permutations of the momenta
$\textbf{p}'_1,\ldots,\textbf{p}'_n$. The tilde reminds that in
general $Tr\tilde{\rho} \neq 1$ and thus, strictly speaking,
$\tilde{\rho}_n$ is not a density matrix. This matrix is generally
used to calculate the $k$-particle momentum distributions
$\Omega_k(\textbf{p}_1,\ldots,\textbf{p}_k)$ putting $n = k$. Thus

\begin{eqnarray}\label{omexcl}
\Omega_1(\textbf{p}) &=& \tilde{C}_1 \rho_I(\textbf{p};\textbf{p}),\\
\Omega_2(\textbf{p}_1,\textbf{p}_2) &=& \tilde{C}_2\left[
\rho_I(\textbf{p}_1;\textbf{p}_1)\rho_I(\textbf{p}_2;\textbf{p}_2)\right.\nonumber
\\ +
\left.\rho_I(\textbf{p}_1;\textbf{p}_2)\rho(\textbf{p}_2;\textbf{p}_1)\right],
\end{eqnarray}
etc., where $\tilde{C}_j$ are normalization constants. This is an
approximation. $\Omega_k$ should have contributions from all $n
\geq k$. A way of taking that into account has been proposed by
Pratt \cite{PRA}. Suppose that without symmetrization the
probability of producing exactly $n$ identical bosons is $P_0(n)$
-- this is another input. Then it is natural to assume that with
symmetrization this probability goes over into

\begin{equation}\label{}
  P(n) = \tilde{C}_0P_0(n)Tr\tilde{\rho}_n,
\end{equation}
where by definition $Tr\tilde{\rho}_0 = 1$. Formulae
(\ref{omexcl}) etc. get replaced by

\begin{eqnarray}\label{omincl}
  \Omega_1(\textbf{p}) = \rho_I(\textbf{p};\textbf{p}) + \nonumber \\\sum_{n=2}^\infty P_0(n)\int dp_2,\ldots,dp_n
  \tilde{\rho}_n(\textbf{p}_1,\ldots,\textbf{p}_n;\textbf{p}_1,\ldots,\textbf{p}_n),
\end{eqnarray}
etc., where the normalization constants have been omitted.
Depending on conventions, $dp$ may mean either $d^3p$ or
$\frac{d^3p}{E_p}$. The difference between formulae (\ref{omexcl})
and analogous and the corresponding formulae (\ref{omincl}) and
analogous is due to the terms with integrations, or in other words
to the multiparticle effects.The problem is (cf. e.g. \cite{ZAJ1},
\cite{HSZ}): how important in practice are these multiparticle
effects?

\section{Low density limit}

Since every permutation can be decomposed into cycles, all the
integrals occurring in Pratt's model can be expressed in terms of
two types of basic integrals. These are the chain integrals:
$G_1(\textbf{p};\textbf{p}') = \rho_I(\textbf{p};\textbf{p}')$ and
for $m > 1$

\begin{eqnarray}\label{}
G_m(\textbf{p},\textbf{p}') = \nonumber \\\int
\rho_I(\textbf{p};\textbf{p}_1)dp_1\rho_I(\textbf{p}_1;\textbf{p}_2)\ldots
dp_{m-1}\rho_I(\textbf{p}_{m-1};\textbf{p}');
\end{eqnarray}
and the cycle integrals

\begin{equation}\label{}
  C_m = \int\! dp\;\; G_m(\textbf{p};\textbf{p}).
\end{equation}
According to this definition $C_1 = 1$. The evaluation of the
basic integrals can be greatly simplified \cite{BIZ1}, \cite{BIZ2}
when the eigenfunctions and eigenvalues of the density matrix
$\rho_I$,

\begin{equation}\label{}
  \int \rho_I(\textbf{p};\textbf{p}')dp'\psi_k(\textbf{p}') = \lambda_k
  \psi_k(\textbf{p}),
\end{equation}
are known. Then

\begin{eqnarray}\label{cmeige}
G_m(\textbf{p};\textbf{p}') &=& \rho_I^m(\textbf{p};\textbf{p}') =\nonumber \\
\label{gmeige} \sum_k \psi_k(\textbf{p}) \lambda_k^m
\psi^*(\textbf{p}'),\\
C_m &=& Tr \rho_I^m = \sum_k \lambda_k^m.
\end{eqnarray}
All the eigenvalues of a density matrix are real, nonnegative and
not exceeding one. Let us label them so that

\begin{equation}\label{}
  0 \leq \ldots \leq \lambda_1 \leq \lambda_0 \leq 1.
\end{equation}
The most populated state has the occupation probability (per
particle) $\lambda_0$. Therefore, the low density limit
corresponds to $\lambda_0 \rightarrow 0$. In this limit all the
basic integrals except $C_1$ and $G_1$ tend to zero and,
consequently, all the integrations drop out. In the low density
limit formula (\ref{omexcl}) and its more-particle analogues
become exact -- the multiparticle effects can be ignored. The
limit of $\lambda_0$ is taken at constant $P_0(n)$. This
corresponds to a constant average number of particles before
symmetrization and, therefore, using the classical terminology,
the density tends to zero, because the amount of available phase
space grows to infinity.

Let us consider an example. Pratt \cite{PRA} chose

\begin{equation}\label{}
  \rho_I(\textbf{p};\textbf{p}') = \frac{1}{\sqrt{2\pi\Delta^2}^3}
  e^{-\frac{(\textbf{p}+\textbf{p}')^2}{8\Delta^2} -
  \frac{1}{2}R^2(\textbf{p}-\textbf{p}')^2}.
\end{equation}
For this choice \cite{BIZ1}, \cite{BIZ2}, \cite{ZAL}

\begin{equation}\label{}
  \lambda_0 = \left(\frac{2}{1 + 2\Delta R}\right)^3.
\end{equation}
The parameter $\Delta^2$  ($R^2$) is equal to the average of the
square of a component of the momentum vector (position vector).
Thus, in the limit $\lambda_0 \rightarrow 0$, which corresponds to
$\Delta R \rightarrow \infty$, the phase space available tends
indeed to infinity.

\section{Finite density}

When $\lambda_0$ does not tend to zero the results become much
more complicated \cite{HSZ}. The two-particle correlation
function, which  according to (\ref{omexcl}) is

\begin{equation}\label{c2excl}
  C_2(\textbf{p}_1,\textbf{p}_2) = 1 +
  \frac{|\rho_I(\textbf{p}_1;\textbf{p}_2)|^2}{\rho_I(\textbf{p}_1;\textbf{p}_1)\rho_I(\textbf{p}_2;\textbf{p}_2)},
\end{equation}
becomes

\begin{equation}\label{c2incl}
  C_2(\textbf{p}_1,\textbf{p}_2) = C_2^{res}(\textbf{p}_1,\textbf{p}_2)\left[ 1 + R_2(\textbf{p}_1,\textbf{p}_2)
  \right],
\end{equation}
where

\begin{eqnarray}\label{}
& C_2^{res}(\textbf{p}_1,\textbf{p}_2) =
\frac{\sum_{i,j}w(i+j)G_i(\textbf{p}_1,\textbf{p}_1)G_j(\textbf{p}_2,\textbf{p}_2)}
{\sum_{i,j}w(i)w(j)G_i(\textbf{p}_1,\textbf{p}_1)G_j(\textbf{p}_2,\textbf{p}_2)},\\
& R_2(\textbf{p}_1,\textbf{p}_2) =
\frac{\sum_{i,j}w(i+j)G_i(\textbf{p}_1,\textbf{p}_2)G_j(\textbf{p}_2,\textbf{p}_1)}
{\sum_{i,j}w(i + j)G_i(\textbf{p}_1,\textbf{p}_1)G_j(\textbf{p}_2,\textbf{p}_2)},\\
& w(m) = \sum_{n=m}^\infty P(n) \frac{n!}{(n-m)!}
\frac{Tr\tilde{\rho}_{n-m}}{Tr\tilde{\rho}_{n}}.
\end{eqnarray}
Function $R_2$ is qualitatively similar to the second term on the
right-hand side of formula (\ref{c2excl}). It is equal one for
$\textbf{p}_1 = \textbf{p}_2$ and tends to zero for $(\textbf{p}_1
- \textbf{p}_2)^2 \rightarrow \infty$. It can be used to extract
information about the geometry of the interaction region. Function
$C_2^{res}$ has been called residual correlation \cite{HSZ},
\cite{ZAJ}. We will see further that it is identically equal one
when the input multiplicity distribution $P_0(n)$ is Poissonian,
but in general it complicates significantly the analysis. In
particular, it can make the value of
$C_2(\textbf{p}_1,\textbf{p}_2)$ at large $(\textbf{p}_1 -
\textbf{p}_2)^2$ bigger than at $\textbf{p}_1 = \textbf{p}_2$
\cite{HSZ}.

Let us consider as an example the case when the input multiplicity
distribution is Poissonian with average multiplicity $\nu$:

\begin{equation}\label{}
  P_0(n) = \frac{\nu^n}{n!} e^{-\nu}.
\end{equation}
Then $w(m) = \nu^m$, which implies $w(i+j) = w(i)w(j)$, and
consequently $C_2^{res}(\textbf{p}_1,\textbf{p}_2) \equiv 1$.
Moreover, introducing the notation \cite{BIZ2}

\begin{equation}\label{}
  L(\textbf{p};\textbf{p}') = \sum_{j=0}^\infty \nu^j
  G_j(\textbf{p};\textbf{p}'),
\end{equation}
one finds \cite{BIZ2}

\begin{equation}\label{}
C_2(\textbf{p}_1,\textbf{p}_2) = 1 +
  \frac{L(\textbf{p}_1;\textbf{p}_2)|^2}{L(\textbf{p}_1;\textbf{p}_1)L(\textbf{p}_2;\textbf{p}_2)}
\end{equation}
which is formula (\ref{c2excl}) with the replacement $\rho_I
\rightarrow L$. Similar relations hold for more-particle
distributions and correlations. This analogy, however, can be
misleading. Interpreting $L(\textbf{p}_1;\textbf{p}_2)$ as if it
were $\rho_I(\textbf{p}_1;\textbf{p}_2)$ one overestimates the
average particle multiplicity $\langle n \rangle$ and
underestimates the averages $\langle \textbf{p}^2 \rangle$ and
$\langle \textbf{x}^2 \rangle$. Thus, the phase space particle
density is overestimated.

\section{How important is all that in practice ?}

The multiparticle effects are negligible when the average
population of each state, or using classical terminology the
particle density in phase space, is sufficiently low. They are
very important when this population, or density, is very high. The
obvious question is: what about the actual data ? We will present
two kinds of estimates suggesting that the multiparticle
corrections are likely to be of the order of 10 \%.

\subsection{Monte Carlo approach}

Most Monte Carlo codes first generate events without Bose-Einstein
correlations. These correlations are introduced later by modifying
the weights of the events or by shifting the momenta of the
generated particles. The former approach has a problem: the
uncorrected Monte Carlo probabilities give multiplicity
distributions in agreement with experiment; the weights modify the
multiplicity distribution and thus tend to spoil its agreement
with experiment. A way out is (cf. e.g. \cite{JAZ}, \cite{FIW}) to
introduce two new parameters: $c$ and $V$ satisfying approximately
the relations

\begin{equation}\label{}
  P_0(n) = c V^{-n}P_{w}(n).
\end{equation}
Here $P_0(n)$ is the uncorrected multiplicity distribution which,
by construction, agrees with experiment. $P_w(n)$ is the
distribution modified by the weight factors. A priori there is no
reason why two parameters should be enough to get reasonable
agreement for the full multiplicity distributions, but in practice
it works. The deviation of the parameter $V$ from unity is a
measure of the importance of symmetrization effects. Very roughly,
it gives the change of the average particle multiplicity as result
of the symmetrization. It was found that $V = 1.1$ for $pp$
scattering at centre of mass energy $\sqrt{s} = 30$ GeV \cite{FIW}
and $V = 1.05$ for $e^+e^-$ annihilations at $\sqrt{s} = 172$ GeV
 \cite{JAZ}. Thus the effects are small, but not
negligible when a precision of some 10 \% is aimed at. These
results are model dependent. It is possible to introduce weights
which oscillate around one so that $V \approx 1$ \cite{KAK},
\cite{KART}. Symmetrization does not affect the multiplicity
distribution at all in the method of momentum shifts \cite{LOS1},
\cite{LOS2}. The shifts affect neither the weight of an event nor
its multiplicity. In this model, however, the authors also find
\cite{LOS2} that multiparticle effects are important, though not
for the multiplicity distribution which is protected by
construction. One important effect of symmetrization, common to
many models, is the enhancement of the single particle momentum
spectra at low momenta.

\subsection{Thermodynamics}

Many models, not only the models called thermodynamic, introduce
the temperature $T$ and the chemical potential $\mu$ of the
particles. In such models the momentum spectrum of the particles
is given before symmetrization by the Maxwell-Boltzmann formula

\begin{equation}\label{}
  \Omega_{MB}(\textbf{p}) = e^{\frac{\mu}{T}}e^{-\frac{E_p}{T}}
\end{equation}
and after symmetrization by the Bose-Einstein formula

\begin{equation}\label{}
  \Omega_{BE}(\textbf{p}) = \frac{1}{e^{\frac{E_p - \mu}{T}}- 1}.
\end{equation}
The proportionality constant in the Maxwell-Boltzmann formula has
been chosen so that at large values of the particle energy $E_p$
the formula coincides with the Bose-Einstein one. The effect of
symmetrization is the "$-1$" in the denominator of the
Bose-Einstein formula. The first remark is that this correction
becomes more important with respect to the exponential when
temperature increases. Incidentally, for this reason some people
consider Einstein's condensation a high temperature effect.

Temperatures in high energy multiple particle production processes
are strongly model dependent. Heavy ion collisions models where
all transverse momentum is due to thermal motion have $T \approx
\langle p_t \rangle \approx 350$ MeV. Models where much of the
transverse momenta is due to collective radial expansion have
lower (local) temperatures of about $100$ MeV. The chemical
potential cannot exceed the lowest possible energy, in this case
$m_\pi = 140$ MeV. Fits assuming temperatures of about $100$ MeV
\cite{TOW}, \cite{AKS} give values of $\mu$ ranging from $30$ MeV
to $90$ MeV. We will give two estimates of the importance of
multiparticle effects  based on these numbers.

The symmetrization effects are the strongest  at $\textbf{p} =
\textbf{0}$. The ratio of the Bose-Einstein to the
Maxwel-Boltznann distribution at $\textbf{p} = \textbf{0}$ is

\begin{equation}\label{}
  r_1 = \frac{1}{1 - e^{-\frac{m_\pi - \mu}{T}}}.
\end{equation}
For $T= 100$ MeV and $\mu$ ranging from $30$ MeV to $90$ MeV $r_1$
ranges from $1.5$ to $2.5$. Thus the correction is very
significant. A more representative ratio is the ratio of the
integrated multiplicities corresponding to the two distributions:

\begin{equation}\label{}
  r_2 = \frac{\int p^2\Omega_{BE}(\textbf{p}) dp}{\int p^2\Omega_{MB}(\textbf{p}) dp}.
\end{equation}
For the same choice of parameters $r_2$ ranges from $1.09$ to
$1.20$, which is consistent with the Monte Carlo results.

\section{Conclusions}

The importance of the multiparticle effects depends both on the
model being used and on the quantities of interest. In Monte Carlo
models, fine tuned to reproduce without symmetrization the
multiplicity distributions, symmetrization is forbidden to affect
the multiplicity distributions. In models related to
thermodynamics the low momentum particles are affected much more
than the high momentum particles. It is not unusual to get from
multiparticle effects corrections of the order of 10 \% or more.
Therefore, it seems that in precision work the multiparticle
effects should not be discarded without previous analysis.

\end{document}